\RequirePackage{amsmath}
\documentclass[epj]{svjour}
\usepackage[english]{babel}
\usepackage{graphics}
\usepackage{amssymb}
\usepackage{graphicx}
\usepackage{amsmath}
\usepackage{dcolumn}
\usepackage{bm}
\usepackage{color}
\usepackage{hyperref}
\usepackage{bookmark}
\usepackage{fancyhdr}
\usepackage{lipsum}
\usepackage{mathtools}
\usepackage{cuted}
\begin{document}
\title{Microscopic Study of ${}^1{S_0}$  Superfluidity in Dilute Neutron Matter}
\titlerunning{Microscopic Study of ${}^1{S_0}$  Superfluidity in DNM}
\author{G. E. Pavlou \inst{1} \thanks{\emph{e-mail:} gepavlou@phys.uoa.gr}, 
E. Mavrommatis \inst{1}, 
Ch. Moustakidis \inst{2}
\and J. W. Clark \inst{3}
%
}                     
%
\institute{Physics Department, Division of Nuclear and Particle Physics, 
National and Kapodistrian University of Athens, GR-15771 Athens, Greece 
\and Department of Theoretical Physics, Aristotelian University of 
Thessaloniki, GR-54124 Thessaloniki, Greece \and McDonnell Center for 
the Space Sciences and Department of Physics, Washington University, 
St. Louis, MO 63130, USA, and Center for Mathematical Sciences, 
University of Madeira, Funchal, 9000-390 Portugal
}
%
\date{Received: {\today} }
%
\abstract{
Singlet $S$-wave superfluidity of dilute neutron matter is studied within the 
correlated BCS method, which takes into account 
both pairing and short-range correlations. First, the equation of state 
(EOS) of normal neutron matter is calculated within the Correlated Basis 
Function (CBF) method in the lowest cluster order using the ${}^1{S_0}$ and 
${}^3P$ components of the Argonne $V_{18}$ potential, 
assuming trial Jastrow-type correlation functions. The ${}^1{S_0}$ 
superfluid gap is then calculated with the corresponding component of the 
Argonne $V_{18}$ potential and the optimally determined correlation 
functions. The dependence of our results on the chosen forms for the 
correlation functions is studied, and the role of the $P$-wave channel 
is investigated. Where comparison is meaningful, the values obtained for 
the ${}^1{S_0}$ gap within this simplified scheme are consistent 
with the results of similar and more elaborate microscopic methods.
\PACS{ {21.65.-f}{Nuclear matter} \and
      {26.60.-c}{Nuclear matter aspects of neutron stars} \and
      {21.60.-n}{Nuclear structure models and methods} \and
      {74.20.Fg}{BCS theory and its development} \and
      {67.10.-j}{Quantum fluids: general properties} 
      } 
} 
%
\maketitle
\section{Introduction}
\label{1.0}

Theoretical study of dilute neutron matter and its superfluid phase continues
to be an active subfield of nuclear theory \cite{Gez14}.  As a model system 
of strongly interacting fermions, it has been a testing ground for advances in 
{\it ab initio} microscopic approaches to quantum many-body problems.
As an essential component of the inner crust of neutron stars, interpenetrating
a lattice of neutron-rich nuclei, dilute neutron matter at baryon densities
in the range $0.2~{\rm fm}^{-1} \lesssim k_F \lesssim 1~{\rm fm}^{-1}$ 
is of considerable importance in dense-matter astrophysics.  Moreover, 
low density neutron matter may also be found in the skins or
outer envelopes of some exotic nuclei.  Additionally, analogies with 
ultracold fermionic atomic gases and their theoretical treatment can be 
fruitfully developed.
\par
The main aim is to study the superfluid dilute neutron matter 
and calculate the pairing gap in the ${}^1{S_0}$ channel. The existence 
of such phase in the inner crust of neutron stars has direct consequences 
for post-glitch relaxation, neutrino emission and cooling of neutron stars 
\cite{Bec09,Haen07,Gan15,Ho15} and an accurate value for the gap is 
required. Many calculations of the ${}^1{S_0}$ superfluid gap have been 
carried out since the 1970s, using various microscopic many-body theories 
(for review see ref.~\cite{Gez14}). Among them we mention direct 
calculations based on the original Bardeen, Cooper, Schrieffer (BCS) 
theory using bare two-nucleon potentials \cite{Kh96,Sch03} and three-nucleon potentials \cite{Mau14,Dri17}, application 
of the polarization potential model of Pines and coworkers \cite{Wam93}, 
inclusion of medium-polarization within a $G$-matrix formulation 
\cite{Sch96,Cao06}, application of Dirac-Brueckner-Hartree-Fock-Bogoliubov 
(DBHFB) theory \cite{Ma08}, calculations using the Correlated Basis 
Function Method (CBF) \cite{Chen93,Fab05,Fab08}, applications of the 
Self-Consistent Green's Function Method \cite{Mu05,Ding16}, a Renormalization 
Group (RG) treatment \cite{Sch03}, and pursuit of various Monte Carlo 
techniques \cite{Gan08,GanC09,Abe09,Gez08,Gez10}.  In spite of these many 
efforts, there is still ambiguity in the value of the ${}^1{S_0}$ gap as 
a function of the density (or $k_F$), owing to the strong sensitivity 
of the gap to inputs for the pairing interaction and self-energies. 
\par
The results reported here were obtained by implementing a generalization
of BCS theory within the CBF framework, giving explicit consideration 
to the role of short-range geometrical correlations induced by the strong 
nuclear force \cite{Krot80,Krot81}. These results supplement previous 
work carried out with the same method \cite{Chen93}, although estimates
of perturbation corrections within the CBF formulation are not included. 
The correlation functions and single - particle energies that enter are 
determined in a straightforward manner from a lowest-cluster order 
calculation of the equation of state (EOS) of normal neutron matter 
including the $S$ and $P$ partial-wave components of the realistic 
Argonne $V_{18}$ potential \cite{Wir95}. 
\par
It turns out that our results for the energy per neutron are, in some 
cases, quite similar to those obtained including higher-order cluster 
contributions and additional components of the Argonne $V_{18}$ potential. 
The EOS of mainly dilute neutron matter has of course been calculated by a broad 
range of many-body techniques (for a review see ref.~\cite{Gan15}).  Among
these are: virial expansion \cite{Ho06}, the lowest-order constrained variational (LOCV) method \cite{Bo98,Bo97,Mo15}, the operator-chain variational 
techniques \cite{Fried81,Akmal98}, CBF calculations \cite{Cl79,Clark79,Bomb05}, 
the Dirac-Brueckner-Hartree-Fock \\ (DBHF) approach \cite{Marg07}, 
Bethe-Brueckner-Goldstone (BBG) methods \cite{Baldo08}, mean field 
calculations with interactions derived by Renormalization Group (RG) 
techniques \cite{Sch03,Heb10}, calculations using inputs from effective 
field theory (EFT) \cite{Sch05,Tew16} and Lattice Chiral EFT \cite{Ep09}.  
Additionally there have been major efforts within the general framework 
of Monte Carlo algorithms, especially variational Monte Carlo (VMC), 
Green's Function Monte Carlo \\ (GFMC) \cite{Gez08,Gez10,Carl03} and 
Auxiliary-Field Diffusion Monte Carlo (AFDMC) \cite{Fab05,Gan08,GanC09,Tew16,Gan09,Lo11}. 
It is expected that the correlation functions that we have determined can 
be used with some reliability in the calculation of other observables of 
low density neutron matter besides the ${}^1{S_0}$ pairing gap. Allowing 
for the simplifications made in our treatment of the ${}^1{S_0}$ gap, the 
results obtained are generally compatible with the estimates reported by other 
authors.  As it is well known, at low densities only the $S$-wave interaction 
is required, whereas at somewhat higher densities of neutron matter, it 
becomes necessary to include the $P$-wave component.  In the density regime 
considered, higher partial waves have negligible impact, and corrections 
from higher-cluster orders and three-nucleon forces are expected to be minimal.

\par
This paper is organized as follows: In sect.~\ref{2.0} we outline our 
method of calculation within the framework of non-orthogonal CBF theory, 
for both the equation of state and the ${}^1{S_0}$ superfluid gap. In 
sect.~\ref{3.0} we present our results and compare them to those of 
other authors.  Section \ref{4.0} is devoted to a brief summary of our results and to some concluding remarks.

\section{Methods of Calculation}
\label{2.0}
\subsection{Equation of State}
\label{2.1}

Our calculations are carried out within the method of correlated
basis functions (CBF) \cite{Cl79}, in which a correlation operator 
${F_N}(1,...,N)$ for $N$ fermions generates not only a trial
ground state wave function but also a complete set of 
non-orthogonal basis states constructed as follows:
\begin{equation}
\label{2.2.1}
{|m \rangle  \equiv |\Psi_m  \rangle  = I_{mm}^{-1/2} F_N(1,\ldots,N)
|\Phi_m \rangle  }.
\end{equation}
Here $\{|\Phi_m \rangle \} $ is a complete orthonormal set of 
states of a suitable independent-particle model and $I_{mm}$ is the 
normalization constant given by
\begin{equation}
\label{2.2.2}
{I_{mm} \equiv  \langle \Phi_m |F_N^\dag(1,\ldots,N)F_N(1,\ldots,N)
|\Phi_m \rangle}.
\end{equation}

To describe normal neutron matter we adopt as model states 
$|\Phi_m \rangle$ a complete orthonormal set of wave functions of 
an ideal Fermi gas of noninteracting neutrons. The Hamiltonian 
matrix elements $H_{mn} = \langle m |\hat H | n \rangle$
in the correlated basis may be used to generate perturbative 
expansions \cite{Cl79} for the ground-state energy and other 
quantities. For the interaction we will consider essential 
components of the Argonne $V_{18}$ two-nucleon potential 
\cite{Wir95}. 

The correlation operator $F_N(1,\ldots,N)$ in eq.~(\ref{2.2.1}) is 
taken as a Jastrow product \cite{Cl79,Clark79} of central two-body 
correlation functions $F_2(ij) = f(r_{ij})$:
\begin{equation}
F_N(1,\ldots,N) = \prod_{i<j} F_2(ij),
\end{equation}
where $r_{ij} = | {\bf r}_i - {\bf r}_j |.$
Denoting the ground state of the noninteracting Fermi gas by 
$|\Phi_0 \rangle$, the Hamiltonian expectation value 
$E_0 = \langle \Psi_0 | H | \Psi_0 \rangle$ 
with respect to the ``ground'' correlated basis state 
\begin{equation}
| \Psi_0 \rangle = \prod_{i<j} F_2(r_{ij}) | \Phi_0 \rangle
\end{equation}
is developed in a cluster expansion \cite{Cl79} in orders of the 
smallness parameter
\begin{equation}
\label{2.2.3b}
\xi  = N^{-1} \sum\limits_{ij} h_{m_im_j,m_im_j},  
\end{equation}
where
\begin{equation}
h_{m_1m_2,m_1m_2}  = \langle m_1m_2 |f^2(r)-1| m_1m_2-m_2m_1 \rangle , 
\end{equation}
with $m_i$ and $m_j$ representing the orbitals of two particles in the 
noninteracting Fermi sea, and $r$ denoting their separation.

For the diagonal matrix elements of the Hamiltonian in the correlated
basis we find
\begin{equation}
\label{2.2.4}
H_{mm}  = \sum\limits_i \varepsilon_{m_i }^{(0)}   
+ \sum\limits_{i < j} w_{m_i m_j,m_i m_j}   + 0(\xi ),
\end{equation}
where the terms $\varepsilon_{m_i}^{(0)}$ are the single - particle 
energies of the orbitals composing the independent-particle model
state $|\Phi_m \rangle$ and 
\begin{equation}
\label{2.2.4a}
w_{m_1 m_2,m_1 m_2} = \langle m_1 m_2 - m_2 m_1 |w_2 (12)| m_1 m_2 \rangle
\end{equation}
are matrix elements of the effective two-body potential which equals
\begin{equation}
\label{2.2.4b}
w_2 (r) = \frac{\hbar^2}{M}\left( {\nabla f(r)} \right)^2 
+ v(r)f^2 (r).
\end{equation}
with the Jastrow correlations.  Here $M$ is the neutron mass 
and $v(r)$ is an appropriate central component of the two-body 
nucleon-nucleon potential. With Fermi-gas energy eigenstates taken for 
the model states $| \Phi_m \rangle$, it is straightforward to
derive the formula
\begin{equation}
\label{2.2.5}
\frac{E}{N} = E_N =  E_F + 2\pi \rho \sum_{\rm S} 
\int_0^\infty w_2^{\rm S}(r)G_{\rm S}({k_F}r) r^2 dr
\end{equation}
for the ground-state energy per particle of neutron matter.
In this expression, (i) $E_F= 3\hbar^2 k_F^2/(10M)$ is the energy per particle for a Fermi gas of non-interacting particles with $M$ the bare neutron mass, (ii) the sum runs over the 
two possible states of two-nucleon spin (singlet ${\rm S}=0$ and 
triplet ${\rm S}=1$), and (iii) $w_2^{\rm S}$ is obtained from 
eq.~(\ref{2.2.4b}) by inserting for $v(r)$ the central potential 
acting in the corresponding two-neutron state of total spin $\rm S$.  
The factors $G_{\rm S}$ are the spatial pair distribution functions in 
singlet and triplet spin states, given by
\begin{equation}
\label{2.2.6}
G_{{\rm S} = 0} (k_F r) = \frac{1}{4}(1+l^2 (k_F r))
\end{equation}
and
\begin{equation}
\label{2.2.7}
G_{{\rm S} = 1} (k_F r) = \frac{3}{4}(1-l^2 (k_F r)),
\end{equation}
where $l(x)$ is the Slater function 
\begin{equation}
\label{2.2.8}
l(x) = 3{x^{-3}}(\sin x - x\cos x).
\end{equation}

Our numerical calculations of the EOS of neutron matter were
based on eq.~(\ref{2.2.5}) along with (\ref{2.2.4b}), (\ref{2.2.6}), and (\ref{2.2.7}).
Ideally, the correlation function $f(r)$ would be determined
by Euler-Lagrange minimization of the ground-state energy expectation
value $E_N$.  Following other authors, we approximate $E_N$ by its
leading cluster order and minimize it with respect to the parameters
of a suitable analytic form for $f(r)$.  This practice has proven 
satisfactory at the low densities involved in the $S$-wave pairing
problem.  In our study, we have considered two forms for $f(r)$, 
(i) the so-called Dav\'e type of ref.~\cite{Chen93}, i.e., 
\begin{equation}
\label{2.2.9}
f_{\rm S}(r) = \exp \left\{{-\frac{1}{2}\left( {\frac{b}{r}} \right)^m 
\exp \left[ {-\left( {\frac{r}{b}} \right)^n } \right]} \right\},
\end{equation}
having parameters $b$, $m$, and $n$, and (ii) the Benhar type \cite{Ben76}
\begin{equation}
\label{2.2.10}
f_{\rm S}(r) = \left[1-\exp\left(-\frac{r^2}{b^2 }\right) \right]^2  
+ gr\exp \left(-\frac{r^2 }{c^2 } \right),
\end{equation}
with parameters $b$, $c$, and $g$.  The parameter $g$ is determined by 
applying the orthogonality condition \cite{Clark79}
\begin{equation}
\label{2.2.11}
\rho \int {d\vec r \left[ {1 - f_{\rm S}(r)} \right]G_{\rm S} (k_F r)}  = 0
\end{equation}
to each spin state where $f(r)$ is written as $f_{\rm S}(r)$ to emphasize that the parameters in the correlation choices \ref{2.2.9} and \ref{2.2.10} can depend on the spin state involved.

Two sets of calculations of the EOS were performed based on the 
Argonne $V_{18}$ interaction.  In the more general case (called 
``spin-dependent''), contributions from both singlet and triplet 
components of the interaction are included.  The singlet component 
is given by its $S$-wave part, i.e., the interaction acting in the 
$^1S_0$ partial wave.  The spin-triplet component, having no $S$-wave 
part in neutron matter, is taken as the interaction acting in the 
$^3P$ partial wave, averaged over the three substates involved, thus
\begin{equation}
\label{2.2.11b}
v_{3P}  = \frac{1}{9}v_{{}^3{P_0}}  + \frac{3}{9}v_{{}^3{P_1}}  
+ \frac{5}{9}v_{{}^3{P_2}}. 
\end{equation}
In the simpler case, only the $^1S_0$ component of the interaction
is included (``$S$-wave only'').

\par
The same two forms of correlation function (Dav{\'{e}} and Benhar types) are 
assumed for ``$S$-wave only'' and ``spin-dependent'' cases, but with 
individually determined parameter values.  These optimal parameters are 
obtained by numerical minimization of the corresponding energy expectation 
value (\ref{2.2.5}) at zeroth order in the small parameter $\xi$ (leading
cluster order), at each density considered.

\subsection{\texorpdfstring{${}^1{S_0}$}{} Superfluid Gap}
\label{2.2}

We adopt Correlated BCS theory \cite{Krot80,Krot81} to study ${}^1{S_0}$ 
superfluidity in neutron matter.  In this theoretical approach, the
non-orthogonal CBF method is used to generalize BCS theory to treat
strongly correlated Fermi systems.  The correlated BCS ground state
is constructed as
\begin{equation}
| {\rm CBCS} \rangle =
\sum\limits_N {\sum\limits_m ( {I_{mm}^{(N)} )}^{-1/2}}
F_N| \Phi_m^{(N)} \rangle \langle \Phi_m^{(N)}
| {\rm BCS} \rangle ,
\label{cbcs}
\end{equation} 
where the kets $| \Phi_m^{(N)} \rangle$ form a complete
orthonormal set of independent-particle (Fermi gas) eigenstates 
and $| {\rm BCS} \rangle$ is the BCS ground state:
\begin{equation}
\label{2.2.13}
| {\rm BCS} \rangle  = \prod\limits_{\vec k} ( {u_{\vec k}} + {\upsilon _{\vec k}}\alpha _{\vec k \uparrow }^\dag \alpha _{ - \vec k \downarrow }^\dag )|0\rangle .
\end{equation}
Here $\vec k$ is the usual wave vector, $u_{\vec k}$ and $\upsilon_{\vec k}$ are 
Bogoliubov amplitudes, and $a_{{\vec k} \uparrow}^\dagger$,
$\alpha_{-{\vec k} \downarrow}^\dagger$ are fermion creation operators, 
with the arrow subscripts indicating spin projections.  
The conventional normalization of (\ref{2.2.13}) to unity 
implies the condition
\begin{equation}
\label{normcond}
u_{\vec k}^2 + \upsilon_{\vec k}^2 = 1  
\end{equation}
on the Bogoliubov amplitudes $u_{\vec k}$ and $v_{\vec k}$. In the normal
state, they reduce respectively to 
${\buildrel_{\circ} \over {u}}{}_{\vec k} =1 - \theta(k_F - k)$ 
and ${\buildrel_{\circ} \over {\mathrm{\upsilon}}}{}_{\vec k}=
\theta(k_F - k)$, where $\theta(x)$ is a step function,
unity for $k_F > k$, zero otherwise and where $k = \left| {\vec k} \right|$.

In the correlated BCS state (\ref{cbcs}), the expectation value of an 
operator ${\hat O}_N$ that conserves the particle number is given by 
\begin{equation}
\label{2.2.16}
{{\left\langle {\hat O} \right\rangle _s} = \frac{{\left\langle {\rm CBCS} 
\right| {{\hat O}_N}\left| {\rm CBCS} \right\rangle }}{{\left\langle 
{{\rm CBCS}} \mathrel{\left | {\vphantom {{CBCS} {CBCS}}} 
\right.\kern-\nulldelimiterspace} {{\rm CBCS}} \right\rangle }}}
\end{equation} 
with the subscript $s$ standing for ``superfluid.''

\par
In analyzing the properties of the correlated BCS state, we employ the commonly 
assumed decoupling approximation \cite{Kh96}, which amounts to treating one 
Cooper pair at a time.  Formally, the ratio (\ref{2.2.16}) is expanded in a 
Taylor series around the normal correlated ground state, retaining terms of 
first order in the deviation of the quantity ${\upsilon_{\vec k}^2}$ from its 
normal-state counterpart ${\buildrel_{\circ}\over{\mathrm{\upsilon}}}{}_{{\vec k}}^2$ 
and of second order in $u_{\vec k}\upsilon_{\vec k}$.  After some algebra \cite{Krot80}, 
one may obtain the following result for the expectation value of 
a number-conserving operator ${\hat O}_N$:

\newpage
\begin{strip}
\begin{equation}
\label{2.2.17}
\begin{split}
{{\langle {\hat O} \rangle }_s} =& O_{oo}^{(A)} \\ & 
+ \sum\limits_{\vec l}^m {( {\upsilon_{\vec l}^2 -{\buildrel_{\circ} 
\over {\mathrm{\upsilon}}}{}_{\vec l}^2} )} \sum\limits_N {\sum\limits_m 
{{{( {I_{mm}^{(N)}} )}^{-1}}} }\prod
\limits_{{\vec k'} \ne {\vec l}}^m {\buildrel_{\circ} \over {\mathrm{\upsilon}}}{}_{{\vec k'}}^2 
\prod\limits_{{\vec k''}}^{\bar m} {\left( {1 - {\buildrel_{\circ} \over 
{\mathrm{\upsilon}}}{}_{{\vec k''}}^2} \right)}\langle 
{\Phi_m^{(N)}} |F_N^\dag ( {{{\hat O}_N} - O_{oo}^{(A)}} )
{F_N} | {\Phi_m^{(N)}} \rangle  \\ & - \sum 
\limits_l^{\bar m} {( {\upsilon_{\vec k}^2 -{\buildrel_{\circ} \over 
{\mathrm{\upsilon}}}{}_{\vec k}^2} )} \sum\limits_N {\sum\limits_m 
{{{( {I_{mm}^{(N)}} )}^{-1}}} }  
\prod\limits_{{\vec k'}}^m {\buildrel_{\circ} \over {\mathrm{\upsilon}}}{}_{{\vec k'}}^2 
\prod\limits_{{\vec k'} \ne {\vec k}}^{\bar m} {( {1 - {\buildrel_{\circ} \over 
{\mathrm{\upsilon}}}{}_{{\vec k''}}^2} )}  \langle 
{\Phi_m^{(N)}} |F_N^\dag ( {{{\hat O}_N} - O_{oo}^{(A)}} )
{F_N} | {\Phi_m^{(N)}} \rangle   \\ & { + \sum
\limits_{\vec l}^m {\sum\limits_{\vec k}^{\bar m} {{u_{\vec l}}{\upsilon_{\vec l}}{u_{\vec k}}{\upsilon_{\vec k}}} } 
\sum\limits_N {\sum\limits_m {{{( {I_{mm}^{(N)}} )}^{ -1}}} }}
{ \prod\limits_{{\vec k'} \ne {\vec l}}^m {\buildrel_{\circ} \over 
{\mathrm{\upsilon}}}{}_{{\vec k'}}^2 \prod\limits_{{\vec k''} \ne {\vec k}}^{\bar m} 
{\buildrel_{\circ} \over {u}}{}_{{\vec k''}}^2 } {  
\langle {\Phi_m^{(N)}} |F_N^\dag ( {{{\hat O}_N} 
- O_{oo}^{(A)}} ){F_N}}{a_{{\vec k} \uparrow }^\dag 
a_{ - {\vec k} \downarrow }^\dag {a_{- {\vec l} \downarrow }}{a_{{\vec l} \uparrow }}
| {\Phi_m^{(N)}} \rangle },
\end{split}
\end{equation} 
\end{strip}

\noindent
where ${\bar m}$ stands for the set of single - particle orbitals
complementary to $m$, the subscript ``oo'' refers to the normal 
correlated state, and $A$ denotes the actual number of fermions. 
Substituting the number operator 
\begin{equation}
\label{noop}
{\hat N = \sum\limits_{{\vec k}} {( {\alpha_{{\vec k} \uparrow }^\dag 
{\alpha_{{\vec k} \uparrow }} + \alpha_{- {\vec k} \downarrow }^\dag 
{\alpha_{- {\vec k} \downarrow }}} )}}
\end{equation}
into this general formula, we find that the number of particles is not 
conserved by the correlated BCS state, as anticipated.  However, number
conservation in the mean is imposed by introduction of a Lagrange 
multiplier $\mu$, identified in general with the chemical potential, and coincident with the single particle energy of the Fermi level $\varepsilon_F$ at zero temperature in the absence of interactions.  Thus, instead of the expectation value 
of $\hat H$ itself, one calculates 
\cite{Krot80,Krot81}
\begin{equation}
\label{2.2.18}
\begin{split}
{\langle {\hat H - \mu \hat N} \rangle _s} =& H_{oo}^{(N)} - \mu N 
+ 2\sum\limits_{{\vec k'} > {{\vec k_F}}} {\upsilon_{{\vec k'}}^2} [{\varepsilon({\vec k'})-\mu} ] \\ 
& {- 2\sum\limits_{{\vec k} < {{\vec k_F}}} {u_{\vec k}^2} [ {\varepsilon ({\vec k}) - \mu } ]} \\ 
& + {\sum\limits_{{\vec k'}} {\sum\limits_{{\vec k} \ne {\vec k'}} {{u_{{\vec k'}}}{\upsilon_{{\vec k'}}}
{u_{\vec k}}{\upsilon_{\vec k}}} } {V_{{\vec k}{\vec k'}}}}
\end{split}
\end{equation}
in terms of the energy $H_{oo}^{(N)}$ of the correlated normal state
of $N$ neutrons, the single - particle energies $\varepsilon ({\vec k})$, and 
the correlated (or effective) pairing matrix elements $V_{{\vec k}{\vec k'}}$.  
\par
The gap function, defined as
\begin{equation}
\label{2.2.19}\Delta_{\vec k} =  -\sum_{{\vec k'}}V_{{\vec k}{\vec k'}}u_{{\vec k'}}\upsilon_{{\vec k'}}
\end{equation}
measures the in-medium binding energy of the Cooper pair.
Applying the Euler-Lagrange variational principle to the form 
(\ref{2.2.18}) while observing the constraint (\ref{normcond})
on the variational Bogoliubov amplitudes, one is led to
the following equation for determination of the gap function in the case of singlet S-wave pairing under consideration:
\cite{Krot80,Krot81}:
\begin{equation}
\label{2.2.20}
\Delta(k) =  - \frac{1}{\pi}\int\limits_0^\infty{\frac{{V(k,k')}}
{{\sqrt{(\varepsilon(k')-\mu)^2  + \Delta ^2 (k')}}}\Delta (k')k'^2 }dk',
\end{equation}
where
\begin{equation}
\label{2.2.21}
V(k,k') = \frac{1}{kk'}\int\limits_0^\infty  
w_2^{{\rm S}=0}(r) \sin (kr)\sin (k'r) dr
\end{equation}
and $w_2^{{\rm S}=0}(r)$ is the effective potential defined by the 
eq.~(\ref{2.2.4b}) for the state ${}^1{S_0}$. The same correlation 
functions are used for both superfluid and normal states.

\par
Equation (\ref{2.2.20}) is a nonlinear integral equation in which
the denominator on the right side becomes vanishingly small as $k'$ 
approaches the Fermi wave number $k_F$.  In spite of this impending 
singularity, straightforward integration of the equation can be 
practical if a suitable starting value of the gap is available 
\cite{Chen93}.  We choose instead to implement the more efficient and 
accurate separation method proposed in ref.~\cite{Kh96}.  Let $\varphi(k) 
= V(k,k_F)/V_F$ and assume that $V_F \equiv V(k_F,k_F) \neq 0$.  The 
matrix elements of the pairing potential are then decomposed identically
as follows
\begin{equation}
\label{2.2.22}
V(k,k') = V_F \varphi(k)\varphi(k') + W(k,k')
\end{equation}
into a separable term and a remainder $W(k,k')$ that vanishes when 
either argument is on the Fermi surface.  Substitution of eq.~(\ref{2.2.22}) 
into the original gap equation (\ref{2.2.20}) leads to an equivalent system 
of two coupled equations for the factors $\chi(k)$ and $\Delta_F$ of
the product $\Delta(k) = \Delta_F \chi(k)$ \cite{Kh96}. The first equation 
is a quasi-linear integral equation for the shape $\chi (k) = 
\frac{{\Delta (k)}}{{{\Delta_F}}}$ of the gap function:
\begin{equation}
\label{2.2.23}
\chi(k) = \varphi (k) 
 - \frac{1}{\pi}\int\limits_0^\infty\frac{W(k,k')\chi(k')k'^2 dk'} 
{\sqrt{(\varepsilon (k') - \mu)^2 + (\Delta_F \chi (k'))^2}}.
\end{equation}
The second equation, which embodies the log singularity, is a nonlinear
integral equation for a number, namely the gap amplitude 
$\Delta_F \equiv \Delta(k_F)$:
\begin{equation}
\label{2.2.24}
1 + \frac{V_F}{\pi}\int\limits_0^\infty\frac{\varphi(k')\chi(k')k'^2 dk'} 
{\sqrt{(\varepsilon (k') - \mu)^2 + (\Delta_F \chi (k'))^2}}=0.
\end{equation}
\par
The first step in finding the gap is evaluation of the pairing matrix
elements of the potential from eq.~(\ref{2.2.21}), using the optimal 
correlation functions determined in the EOS calculation outlined in
sect. \ref{2.1}.   For the single - particle energies, we adopt an 
effective-mass approximation i.e., $\varepsilon(k) = \hbar^2 k^2/2M^* 
+ {\rm const.}$, which should be satisfactory at the low densities
in question.  Specifically, with $\varrho(k) \equiv d\varepsilon(k)/dk$, 
$M^*$ is estimated as
\begin{equation}
\label{2.2.25} 
{M^*} = \hbar^2 k_F \varrho^{-1}(k_F).
\end{equation}
In the general case where the singlet-$S$ and triplet-$P$ components
of the potential are included, the single - particle energies are expressed
more explicitly as
\begin{equation}
\label{2.2.26}
\begin{split}
\varepsilon (k) = \frac{{{\hbar ^2}{k^2}}}{{2M}} 
& + \pi \rho \int\limits_0^\infty {{r^2}w_2^{\textrm{S}=0}(r)
\left[ {1 + \frac{{\sin (kr)}}{{kr}}l(kr)} \right]dr}  \\ 
& + 3\pi \rho \int\limits_0^\infty {{r^2}w_2^{\textrm{S}=1}(r)\left[ {1 - \frac{{\sin (kr)}}{{kr}}l(kr)} \right]dr.} 
\end{split}
\end{equation}

Given these preparations, eqs.~(\ref{2.2.23}) and (\ref{2.2.24}) are
solved by iteration starting from a constant value for ${\Delta_F\chi(k')}$ 
in eq.~(\ref{2.2.24}), continuing until satisfactory 
convergence is achieved for ${\Delta _F}$ \cite{Kh96} -- typically in
very few steps.

\section{Results and Discussion}
\label{3.0}
\subsection{Equation of State}
\label{3.1}

\begin{figure}
\centering
\resizebox{0.5\textwidth}{!}{%
\includegraphics{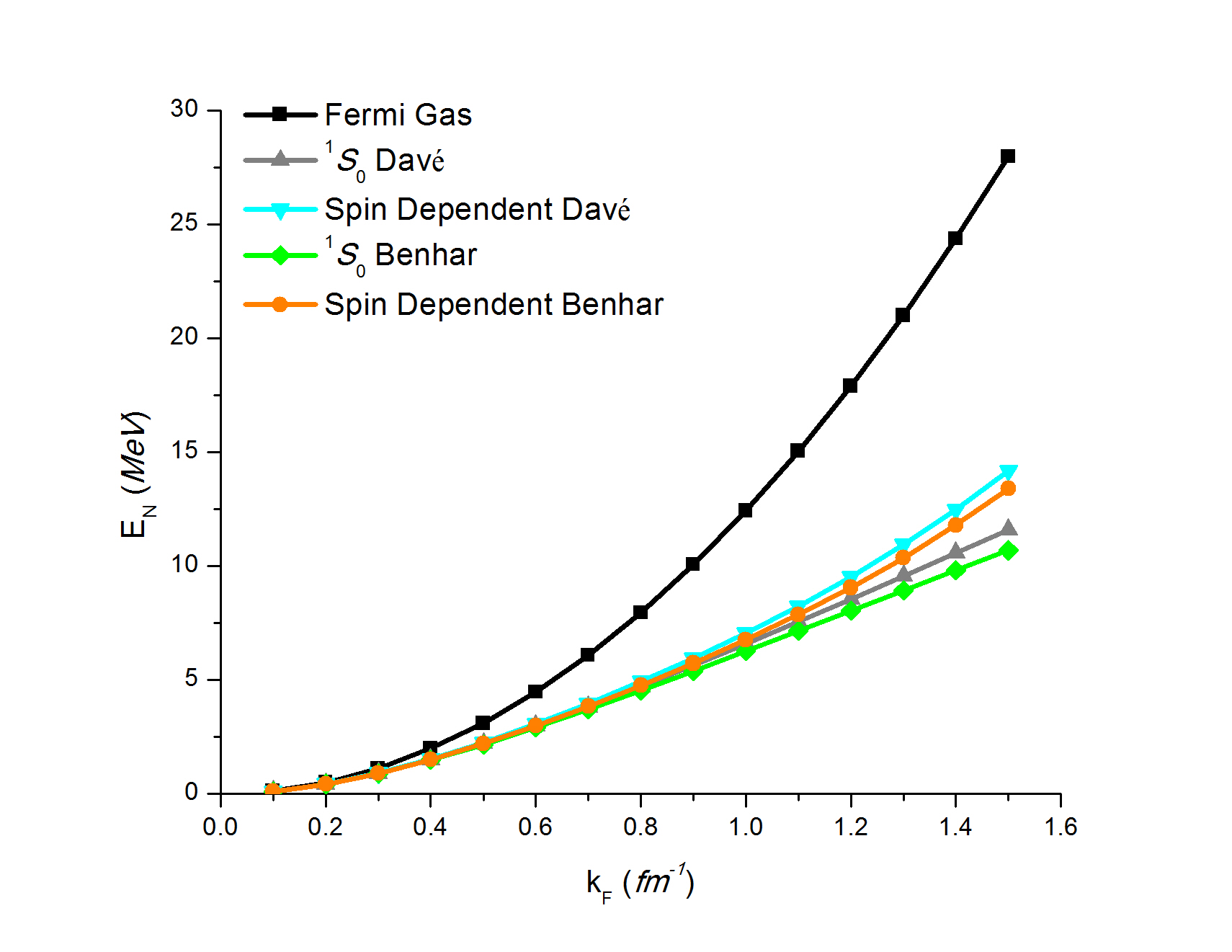}
}
\vspace{-0.5cm}
\caption{(Color online) Energy per neutron $E_N$ as a function of the 
wave number $k_F$ based on the Argonne $V_{18}$ interaction in the four 
calculational treatments as described in the text.  The energy per 
neutron of the corresponding ideal Fermi gas is plotted for comparison.}
\label{fig:U18}
\end{figure}

\begin{figure}
\centering
\resizebox{0.45\textwidth}{!}{%
\includegraphics{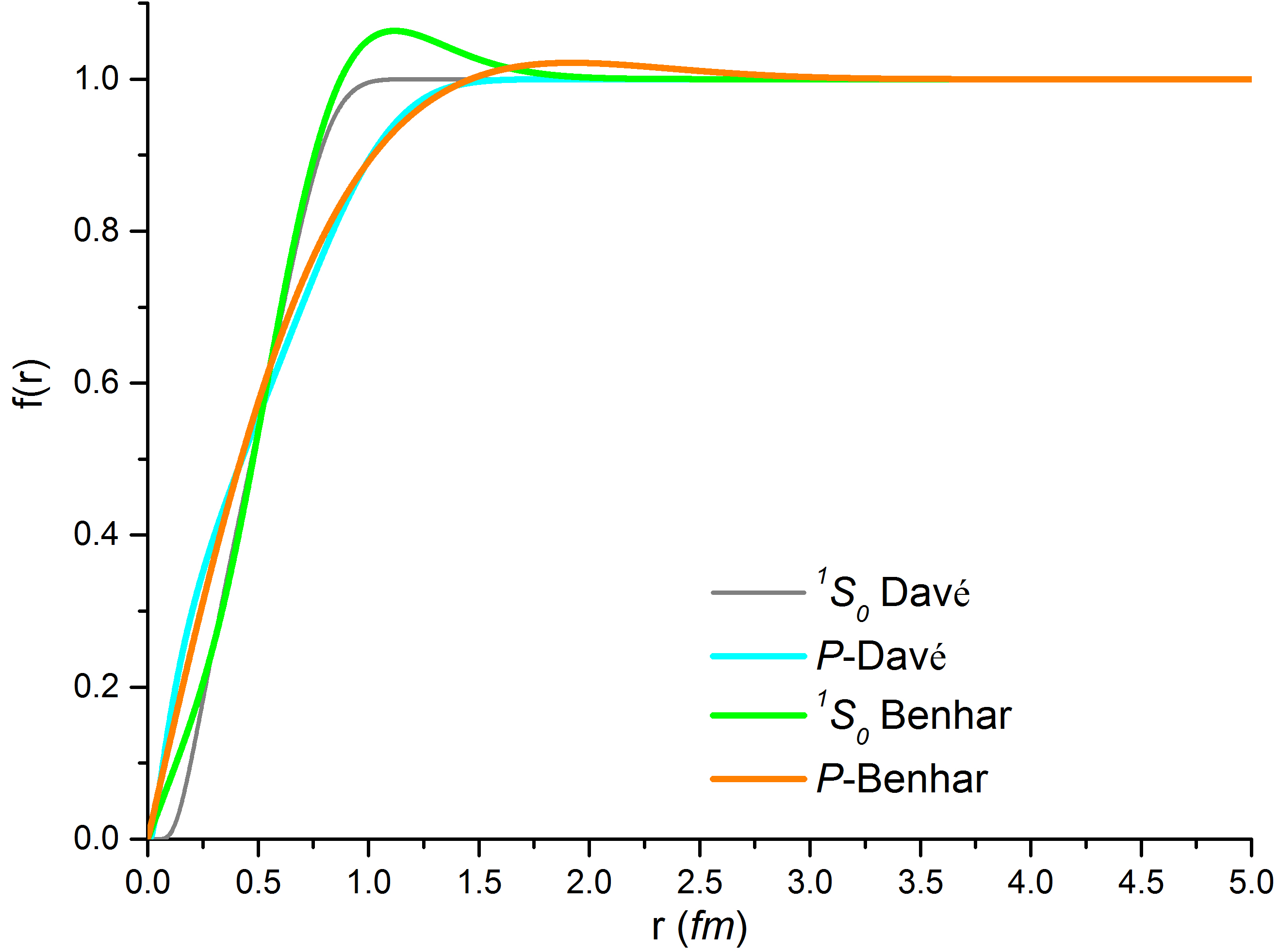}
}
\vspace{-0.0cm}
\caption{(Color online) Correlation functions of the Jastrow type 
$f_{\rm S}(r)$ (Dav\'e and Benhar versions), determined optimally at Fermi wave 
number $k_F=0.9~{\rm fm}^{-1}$ and plotted versus the separation $r$ 
of a pair of neutrons interacting (i) via the $S$-wave component of 
the Argonne $V_{18}$ potential (``$S$-wave only''), and (ii) also via the $P$ component of this potential.}
\label{fig:correlation}
\end{figure}

We first present the results for the EOS of normal dilute neutron matter. 
In fig. \ref{fig:U18}, the optimized energy per neutron is shown as a function of the Fermi wave number $k_F$, for each of the four cases 
considered for the Argonne $V_{18}$ two-nucleon interaction ($S$-waves only;
$S$ and $P$ waves; Dav\'e and Benhar correlation functions).  
Figure~\ref{fig:correlation} features plots of the optimized $S$ and $P$ 
correlation functions $f_{\rm S}(r)$ at the typical value 
$k_F = 0.9~{\rm fm}^{-1}$, for each of the two types considered.  Almost identical 
results \cite{Pavlou09} were obtained with the corresponding components 
of the Argonne $V_{4'}$ potential \cite{Wir02}.  Even though $k_F$ 
values only up to about $1~{\rm fm}^{-1}$ are needed to describe the 
inner crust of neutron stars, the EOS has been plotted for $k_F$ up 
to $1.5~{\rm fm}^{-1}$, in order to show the influence of the $P$-state 
contribution as the density increases.

\par
Individually for the ``$S$-wave only'' and ``spin-dependent'' cases,
the energetic results obtained for the optimal Dav\'e and Benhar correlation
functions are found to be rather close, but with a discrepancy that 
increases with density.  This is quite as expected, since the differences 
between these functions at short range is better resolved at higher 
densities.  Further, it is to be noted that the Dav\'e form has three 
free parameters, compared to two in the Benhar case.  Another distinction between these 
two correlation functions, seen in fig.~\ref{fig:correlation}, is that 
the Benhar form allows $f_{\rm S}(r)$ to overshoot unity at small $r$, whereas 
the Dav\'e form does not.  Comparing the energetic results obtained 
in the ``$S$-wave only'' and ``spin-dependent'' cases for a given correlation 
form, it is seen that inclusion of the positive $P$-wave contribution 
begins to play a role with increasing density, such that its incorporation 
becomes necessary for $k_F$ values beyond about $0.8~{\rm fm}^{-1}$.

Of special interest is the magnitude of the ``smallness parameter''
$\xi$ defined in eq.~(\ref{2.2.3b}), which governs the convergence
of the cluster expansion of the energy.  Over the densities and 
correlation functions considered, this parameter rises monotonically with the density and ranges from
$5.66 \times 10^{-6}$ for $k_F =0.1~{\rm fm}^{-1}$ to $0.006$ for $k_F = 1~{\rm fm}^{-1}$ and to $0.026$ for 
$k_F = 1.5~{\rm fm}^{-1}$ for the Benhar case and from $5.36 \times 10^{-6}$ for $k_F =0.1~{\rm fm}^{-1}$ to $0.006$ for $k_F = 1~{\rm fm}^{-1}$ and to $0.027$ for 
$k_F = 1.5~{\rm fm}^{-1}$ for the Dav\'e case. On this basis, the higher cluster corrections
neglected in the present study should not be important below
about $k_F = 1.0~{\rm fm}^{-1}$, as far as the EOS itself is concerned.

\begin{figure}
\centering
\resizebox{0.5\textwidth}{!}{%
  \includegraphics{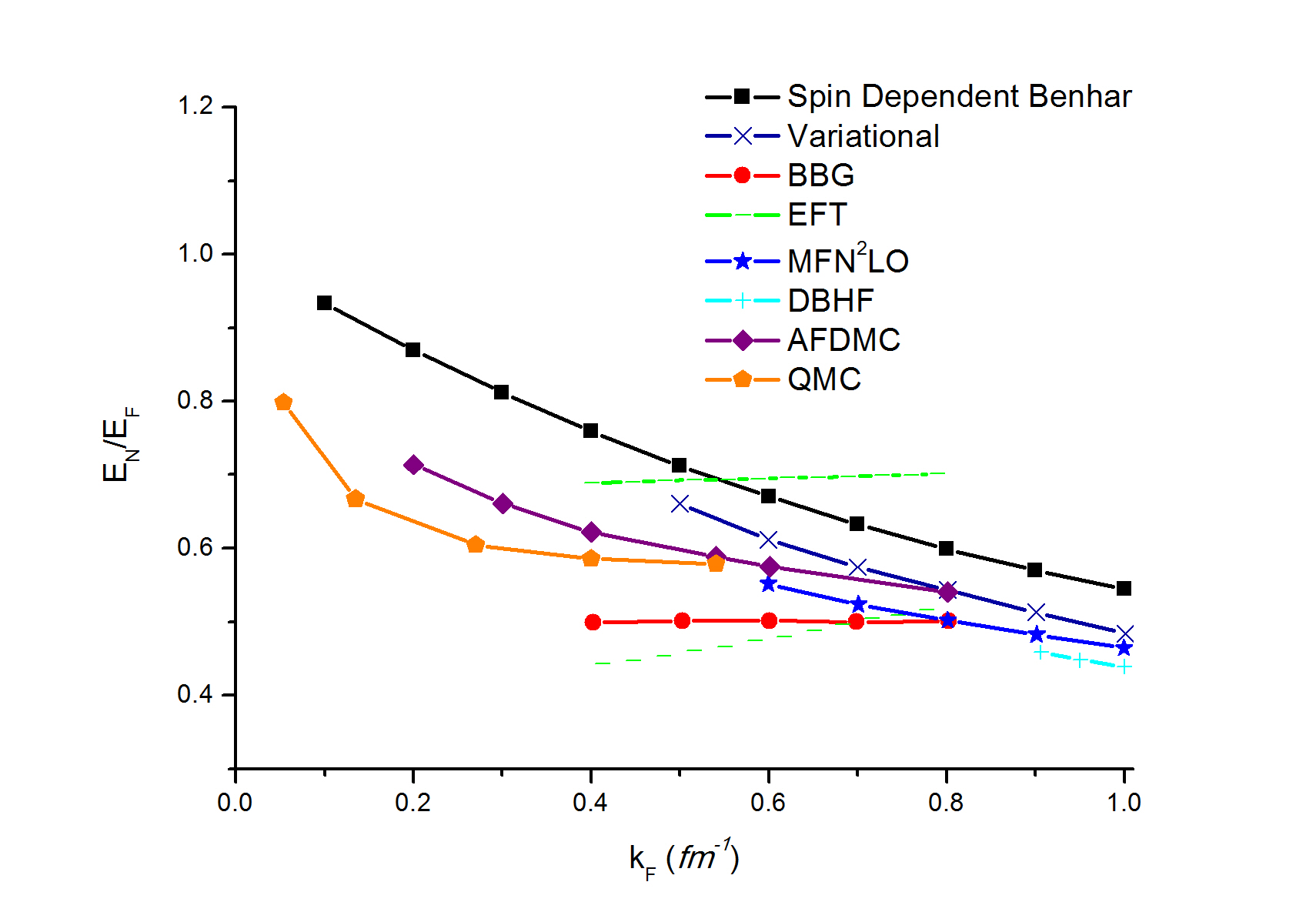}
}
\vspace{-0.7cm}
\caption{(Color online) Comparison of results for the ratio of the energy per
neutron $E_N$ of neutron matter (i.e., its EOS) to the energy per particle 
$E_F$ of a Fermi gas of noninteracting particles of the same mass,
obtained by different microscopic many-body methods, generally with differing
input potentials, considered to be realistic.  See text for details.
The curve marked with black squares traces the EOS obtained here for 
the Argonne $V_{18}$ interaction when the Benhar correlation function 
is adopted in the ``spin-dependent'' case.}
\label{fig:EOS}
\end{figure}

\par
Results obtained for $E_N$ with the Benhar correlation function 
for the Argonne $V_{18}$ interaction in the ``spin-dependent'' case are 
plotted in fig.~\ref{fig:EOS} along with results from other 
computational many-body methods. In choosing between the Dav\'e or Benhar forms of the correlation function $f_{\rm S}(r)$ to show results on fig. \ref{fig:EOS} (and later on fig. \ref{fig:delta}) we have decided in favor of the former for two reasons.  First, it gives a slightly lower upper bound for the energy. The second reason concerns the question whether $f_{\rm S}(r)$ should overshoot unity before going to zero to suppress the core region of the potential.  There are some studies that support the presence of such an overshoot:  (i) LOCV calculations of the energy per particle (see for example refs. \cite{Mo15}), which give results close to those of FHNC/SOC and MC calculations upon solving Euler-Lagrange equations at the two-body cluster level, produce correlation functions that seem to overshoot unity.  (ii) Recently, optimized FHNC indicate a small overshoot in neutron matter for the Argonne $V_{18}$ interaction \cite{private}. The methods shown in fig. \ref{fig:EOS} include Variational approaches that introduce state-dependent correlations 
\cite{Bo97,Bo98,Mo15,Fried81}, \\ Brueckner-Bethe-Goldstone (BBG) theory \cite{Baldo08}, 
difermion effective field theory (EFT) \cite{Sch05}, a Mean Field 
calculation using chiral ${\rm N}^2$LO three-nucleon forces 
(MF${\rm N}^2$LO) (showing error bounds) \cite{Heb10}, 
Dirac-Brueckner Hartree-Fock theory (DBHF) \cite{Marg07}, 
Auxiliary Field Diffusion Monte Carlo (AFDMC) \cite{Gan09} 
(including a three-nucleon interaction), and Quantum Monte Carlo 
(QMC) \cite{Gez08,Gez10}.  The most meaningful comparison of the
present results would be with the variational calculations of 
Friedman and Pandharipande (FP) \cite{Fried81} and of Modarres et. al. \cite{Bo97,Bo98,Mo15}. It is understandable 
that the ``spin-dependent'' Benhar curve would lie somewhat above that 
of FP, since the latter calculation involves a more flexible 
variational ansatz and includes essentially all components 
of the assumed two-nucleon interaction. The same is true for the results obtained with the LOCV method which lie very near the ones of the FP paper, but are available for $k_F$'s $\gtrsim 1f{m^{ - 1}}$. Results of ref. \cite{Mo15} are below those of refs. \cite{Carl03} and \cite{Lo11} obtained with GFMC and FHNC methods respectively. It is worth noting at this
point that at similar densities, precise calculation of the EOS of pure 
neutron matter is less demanding than that for symmetrical nuclear matter, 
where large-scale cancellations occur between kinetic and potential
contributions to $E_N$.  An additional consideration is that in
the density regime of neutron matter of interest for $^1S_0$ 
pairing, three-nucleon interactions are not expected to play
a very significant role \cite{Mau14,Dri17,Ham13,zuo,zhou}.

\subsection{\texorpdfstring{${}^1{S_0}$}{} Superfluid Gap}
\label{3.2}

\begin{figure}
\centering
\resizebox{0.5\textwidth}{!}{%
  \includegraphics{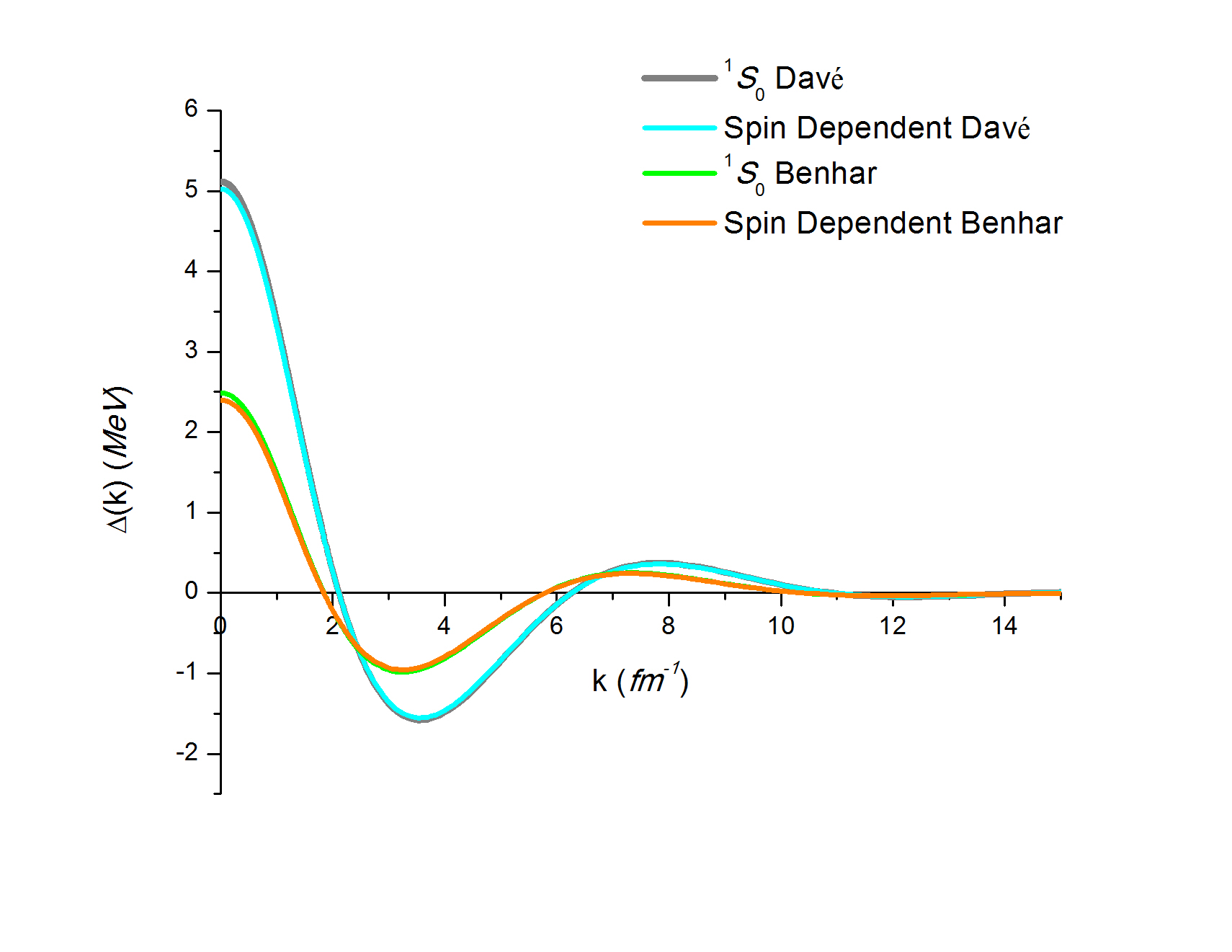}
}
\vspace{-10mm}
\caption{(Color online)  Gap function $\Delta(k)$ at $k_F=0.9~{\rm fm}^{-1}$ 
as a function of wave number $k$, obtained with an effective pairing 
interaction determined from eq.~(\ref{2.2.4b}), based on the Argonne
$V_{18}$ interaction.  The four curves correspond to the four different 
choices for the Jastrow correlation function $f_{\rm S=0}(r)$, as identified in 
the main text.}

\label{fig:gapfunc}
\end{figure}

As explained above, we solve the gap equation (\ref{2.2.20}) using 
the separation method of Khodel, Khodel, and Clark \cite{Kh96} 
and the optimal correlation functions $f_{\rm S=0}(r)$ determined here for the normal 
neutron matter. In fig.~\ref{fig:gapfunc}, the resulting gap function
$\Delta (k)$ is plotted as a function of $k$ for $k_F =0.9~{\rm fm}^{-1}$ 
for all four correlation choices $f_{\rm S=0}(r)$ ("S-wave only", "Spin Dependent" Dav\'e and Benhar forms) based on the Argonne $V_{18}$ 
interaction.  An interesting feature prominent in these plots is the 
occurrence of a node in the gap function at $k \simeq 2~{\rm fm}^{-1}$, 
which is generic to pairing interactions that possess a substantial inner 
repulsion in coordinate space, along with the outer attraction required by
the experimental $^1S_0$ phase shift.  Due to the non-monotic behavior 
of the interaction, the negative excursion of the gap function is
generally necessary for the existence of a solution of the gap equation,
as emphasized in ref.~\cite{Kh96}.  A feature specific to the present 
calculation is the tiny discrepancy between the gap functions generated 
in the $S$-wave only and ``spin-dependent'' cases.  The effect of the 
$P$-wave component of the two-nucleon potential on the optimal 
correlation function $f_{\rm S=0}(r)$ of either type is minuscule in the gap function.

\begin{figure}
\centering
\resizebox{0.5\textwidth}{!}{%
  \includegraphics{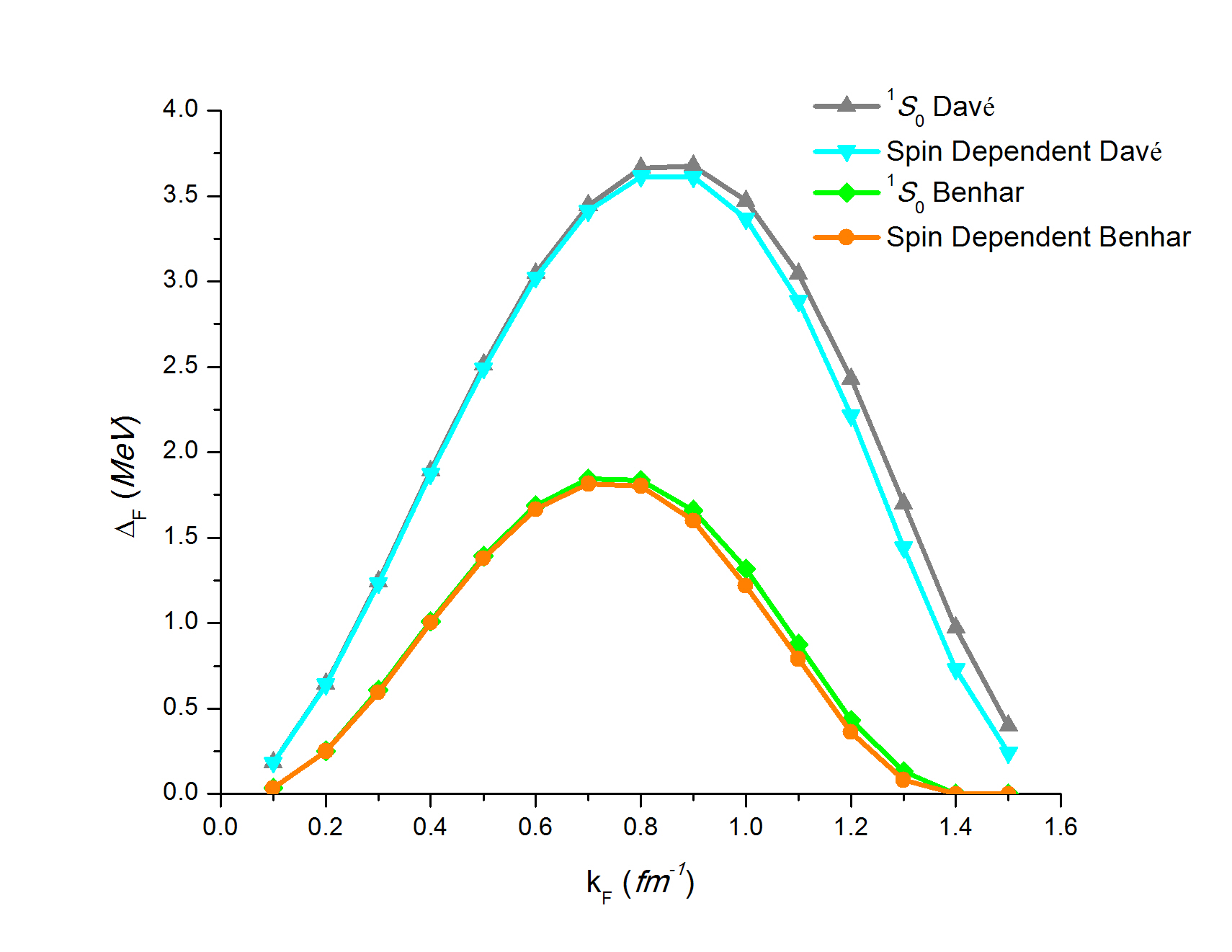}
}
\vspace{-0.7cm}
\caption{(Color online) Pairing gap $\Delta_F$ on the Fermi surface 
for the ${}^1{S_0}$ superfluid state of neutron matter as a function 
of Fermi wave number $k_F$, as obtained with an effective pairing 
interaction determined from eq.~(\ref{2.2.4b}) based on the Argonne
$V_{18}$ interaction.  
The four curves correspond to the four different choices for the 
Jastrow correlation function $f_{\rm S=0}(r)$, as identified in the main text.
}
\label{fig:gap}
\end{figure}

The resulting energy gap on the Fermi surface, $\Delta(k_F) \equiv 
\Delta_F$, is plotted in fig.~\ref{fig:gap} as a function of $k_F$ 
for the case of the Argonne $V_{18}$ interaction. As a check, we have 
also solved the gap equation by straightforward iteration, taking 
proper care in dealing with the small values of the denominator around 
the Fermi surface.  Good agreement was found \cite{Pavlou09}, the essential
point being that the Jastrow correlations act to ``tame'' the extreme 
non-monotonicity of the bare interaction.

\par
Comparing the two parametrized
forms for the correlation function we realize sensitivity. In the Dav\'e case we found a
somewhat larger range of $k_F$ values over which a non-zero gap
exists, than for the Benhar form, as well as a somewhat larger
range of gap values.  Comparing the gap values in the $S$-wave-only
case with those for which the $P$-wave contribution affects the
optimal choice of the correlation function $f_{\rm S=0}(r)$, we find that the latter are slightly 
smaller than the former for $k_F \gtrsim 0.7~{\rm fm}^{-1}$. 
Accordingly, we again affirm that inclusion of the $P$ channel in
our procedure has a small negative effect on the gap. This 
is true for either form chosen for the correlation function.  
Concerning the Fermi wave number (effectively the density) at which
the peak value $\Delta_F$ is reached, it is located between 
$k_F=0.8~{\rm fm}^{-1}$ and $k_F=0.9~{\rm fm}^{-1}$, for the Dav\'e 
correlations, while for Benhar correlations it lies between 
$k_F=0.7~{\rm fm}^{-1}$ and ${k_F} = 0.8~{\rm fm}^{-1}$.  With 
respect to the (upper) density at which the gap closes, our approach 
gives a values close to $k_F = 1.5~{\rm fm}^{-1}$ and $1.3~{\rm fm}^{-1}$ 
for Dav\'e and Benhar choices, respectively. While differences between Benhar and Dav\'e forms may not matter much at all for the energy, the different balance between the positive and negative parts of the effective pairing interactions for these two choices may matter a great deal in determining the gap, the more so when compounded with the quasi-exponential dependence of this quantity on the inputs for the pairing interaction and self-energy.  Here it is again prudent
to mention that although $k_F$ does not exceed about $1~{\rm fm}^{-1}$
in the neutron-star inner crust, we have chosen to plot the gap results
up to $1.5~{\rm fm}^{-1}$ in order to show how the $P$-wave component
of the two-nucleon interaction might affect, through the two-body 
correlations it influences, the density at which the gap closes.
Analogous calculations have been performed for the Argonne $V_{4'}$ potential,
with results \cite{Pavlou09} that show very little difference from 
those reported here for the Argonne $V_{18}$ interaction.

\begin{figure}
\centering
\resizebox{0.5\textwidth}{!}{%
\includegraphics{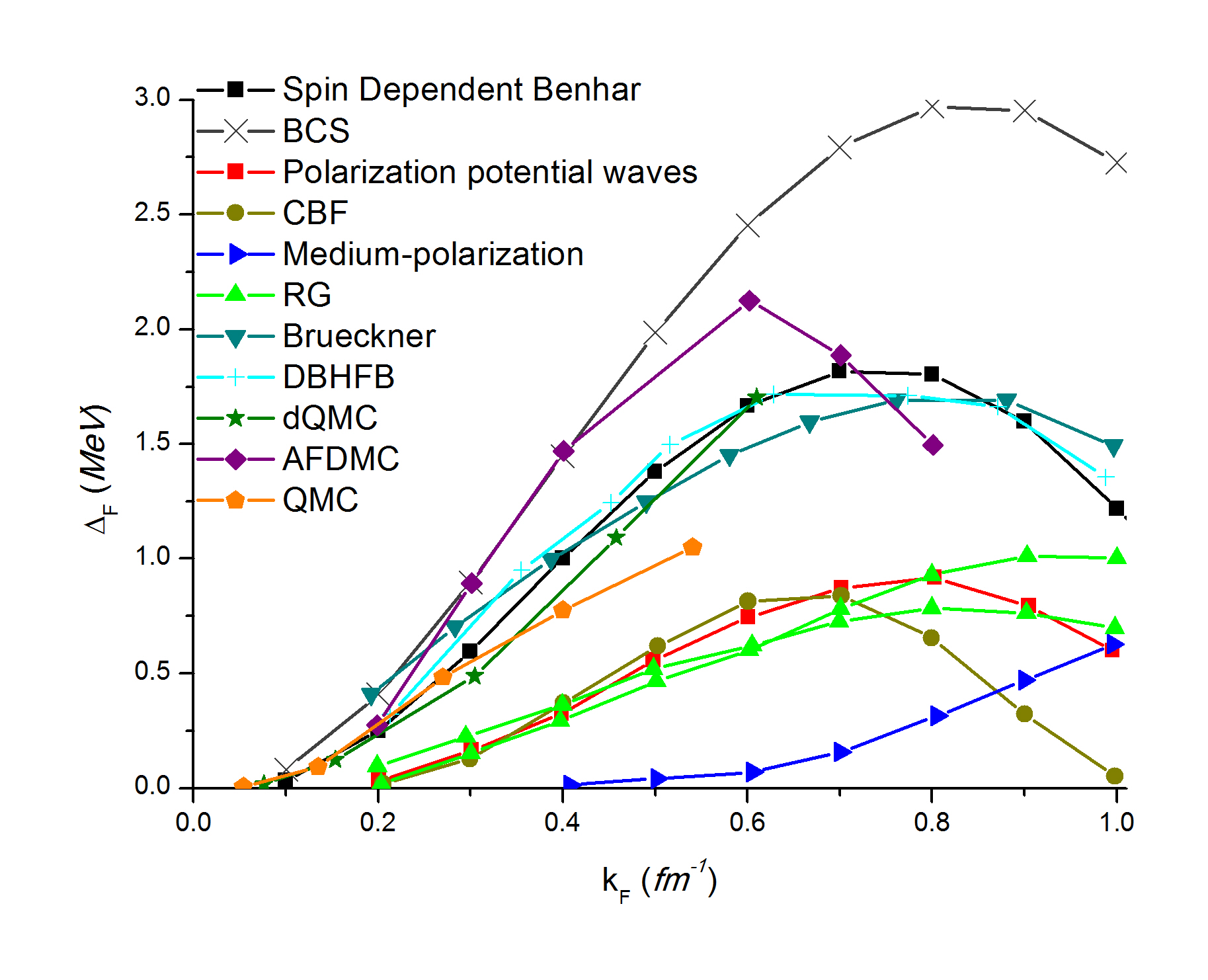}
}
\vspace{-0.8cm}
\caption{(Color online) Results for the ${}^1{S_0}$ neutron gap 
$\Delta_F$ versus Fermi wave number $k_F$, as obtained for 
the Argonne $V_{18}$ interaction using the Benhar-type correlation 
function in the ``spin-dependent'' case (curve marked with black squares).  Results for $\Delta_F$
calculated by other microscopic methods are displayed for comparison
(see text for details).}
\label{fig:delta}
\end{figure}

\par
Our results for the $S$-wave gap may be compared with those of 
refs.~\cite{Kh96,Pavlou09} that were obtained from ordinary BCS theory 
using the bare $^1S_0$ component of the $AV_{18}$ potential as pairing 
interaction (thus including no corrections for short-range geometrical 
correlations or medium polarization). The corresponding curve 
of $\Delta_F$ versus $k_F$ is marked with crosses in fig.~\ref{fig:delta}. 
The gap values calculated with Benhar-type correlations in the 
``spin-dependent'' case are seen to be suppressed relative to the 
pure BCS gap by a factor of about 2/3 in region of the peak, which 
occurs at a slightly lower density in our calculation.

\par
Since the Dav\'e form for the Jastrow correlation function was employed 
in the non-orthogonal CBF approach applied by Chen et al.~\cite{Chen93} to the problem of $^1S_0$ neutron pairing, it is of special interest to compare 
the results obtained here with the Dav\'e form (``spin-dependent'' case) 
with those from this earlier CBF calculation (plotted as ``CBF'' in 
fig.~\ref{fig:delta}).  Other than in the methods used to solve the gap 
equation, both of which are sufficiently accurate, the difference 
between the two studies lies primarily in the inclusion, by Chen et al., 
of a correction of second order in CBF perturbation theory to account 
for in-medium modification of the pairing interaction (``polarization 
effects''). However this correction is very approximately estimated and somewhat questionable, as discussed in refs. \cite{Chen93} and, more recently, in \cite{world}.  Absent that correction, the results for the gap are found 
to be very similar, as expected.  The impact of differences in the 
pairing interactions assumed (Reid $V_4$ in ref.~\cite{Chen93} and 
Argonne $V_{18}$ herein) is minimal.

Also reproduced in fig.~\ref{fig:delta} are gap results obtained through
a number of other microscopic many-body methods, diverse in
their inclusion (or not) of various physical effects influencing
the pairing gap.  These methods include the polarization-potential 
model of Pines et al. \cite{Wam93}, a medium-polarization 
calculation \cite{Sch96}, an application of RG theory \cite{Sch03}, 
approaches grounded in Brueckner theory \cite{Cao06}, DBHFB \cite{Ma08}, 
determinantal lattice QMC (dQMC) \cite{Abe09}, AFDMC \cite{Gan08}, 
and QMC \cite{Gez10}.  Other noteworthy calculations not represented 
in fig. \ref{fig:delta} are those of refs.~\cite{Mau14,Dri17,Fab05,Fab08,Mu05,Ding16}.

\par
Qualitatively, our results for $\Delta_F$ using Benhar correlations 
are closer to those of refs.~\cite{Cao06,Ma08,Abe09,Gez08,Gez10}, while
differing substantially from those of refs.~\cite{Sch03,Wam93,Sch96,Chen93}. 
The density at which our predicted gap reaches a maximum is similar 
to what found in refs.~\cite{Wam93,Sch96,Cao06,Ma08}. However, useful
conclusions cannot be drawn from such commonality or disparity, due
to differences in the pairing interactions assumed and in calculational
methods, especially the treatment (or not) of in-medium modification
of the bare interaction.

It is not surprising that considerable uncertainty remains in the 
quantitative determination of the behavior of the $^1S_0$ gap 
$\Delta_F$ in neutron matter, in view of the inherent strong
sensitivity to the inputs for the pairing interaction and the
self-energies (or density of states).

\section{Summary and Conclusions}
\label{4.0}

The ${}^1{S_0}$ superfluid gap $\Delta_F$ of dilute neutron matter in 
the density range $0.2~{\rm fm}^{-1} \lesssim k_F \lesssim 1~{\rm fm}^{-1}$ 
has been calculated in the framework of correlated BCS theory at lowest 
cluster order, an approximation suitable in the low-density regime where 
pairing occurs in the singlet $S$-wave state of two neutrons.  
Inputs to this theory consist of the bare Argonne $V_{18}$ 
nucleon-nucleon interaction and corresponding optimized Jastrow-type 
correlation functions that modify it to create an effective pairing 
interaction that takes account of the effects of strong short-range 
correlations present in nuclear systems.  The CBF gap equation has
been solved for this effective pairing interaction using the 
robust and accurate separation method introduced in ref.~\cite{Kh96}.  

A many-body approach limited to inclusion of the effects of $S$- 
and $P$-wave components of the $AV_{18}$ interaction is adequate at 
the low densities relevant to pairing in the $^1S_0$ state.  The 
parameters of the chosen forms of Jastrow correlation function $f(r)$ 
have been determined by minimization of the expectation value of 
the system Hamiltonian with respect to the Jastrow trial function

\begin{equation}
\label{2.4.1}{| \Psi_0 \rangle = \prod_{i<j} f(r_{ij}) |\Phi_0 \rangle,}
\end{equation}
\noindent
where $|\Phi_0 \rangle$ is the ground state of the neutron system 
with interactions turned off.  In practice, this expected energy 
is evaluated to leading order in a small parameter $\xi$.
Roughly speaking, this parameter is given by the ratio of the 
volume per particle in which $f^2(r)-1$ is appreciable, to the 
total volume per particle. Thus, the many-body description adopted 
rests on a low-density approximation, presumed to be adequate for 
description of the low-density neutron matter in the inner crust 
of a neutron star, where, roughly, $\xi \sim 0.1$.  As a by-product, 
execution of this optimization process yields a corresponding 
approximation to the ground-state energy per particle of neutron 
matter, i.e., its equation of state (EOS), in the relevant density 
regime ranging up to about one-fourth the saturation density of 
symmetrical nuclear matter.  These EOS results have been contrasted 
with those generated by other many-body methods, generally with 
other choices of basic interactions.  In our calculation, the 
contributions of the $P$-wave and higher partial-wave channels 
become important for the EOS, and indirectly to the $^1S_0$ pairing gap, 
only for $k_F \gtrsim 0.8~{\rm fm}^{-1}$, as must also be the case in
other studies of neutron matter in this density regime.  Similarly, 
fundamental three-nucleon forces should have only modest impact on the 
EOS and $^1S_0$ superfluid gap in this regime \cite{Mau14,Dri17,Ham13,zuo,zhou}. 

Comparison of the results for the $^1S_0$ gap obtained here with
the results of the numerous antecedent calculations is obscured by the 
wide range of both many-body methods employed and types of input 
interactions assumed.  One significant finding is a quenching
of the gap obtained with the effective pairing interaction 
generated by the CBF approach, relative to the gap predicted by
pure BCS theory for the bare Argonne $V_{18}$ interaction. It 
may also be noted that the gaps $\Delta_F$ obtained with our approach 
are found to be rather close to those calculated by Gezerlis 
and Carlson \cite{Gez10} using quantum Monte Carlo techniques
(also for the Argonne $V_{18}$ interaction), and similar to results
from a determinantal lattice QMC approach \cite{Abe09}.

%
 \bibliographystyle{epj}
 \bibliography{mybibv3}

\begin{thebibliography}{52}

\bibitem{Gez14}
A.~Gezerlis, C.J. Pethick, A.~Schwenk, \emph{{Novel Superfluids: Vol. 2
  \textnormal{Chapter 22, edited by K. H. Bennemann and J. B. Ketterson}}}
  (Oxford University Press, 2014)

\bibitem{Bec09}
W.~Becker, ed., \emph{{Neutron Stars and Pulsars}}, Vol. 357 of
  \emph{Astrophysics and Space Science Library} (Springer-Verlag Berlin
  Heidelberg, 2009)

\bibitem{Haen07}
P.~Haensel, A.Y. Potekhin, D.G. Yakovlev, eds., \emph{{Neutron Stars 1 :
  Equation of State and Structure}}, Vol. 326 of \emph{Astrophysics and Space
  Science Library} (Springer-Verlag New York, 2007)

\bibitem{Gan15}
S.~Gandolfi, A.~Gezerlis, J.~Carlson, Annu. Rev. Nucl. Part. Sci. \textbf{65},
  303 (2015)

\bibitem{Ho15}
W.C.G. Ho, K.G. Elshamouty, C.O. Heinke, A.Y. Potekhin, Phys. Rev. C
  \textbf{91}, 015806 (2015)

\bibitem{Kh96}
V.A. Khodel, V.V. Khodel, J.W. Clark, Nucl. Phys. A \textbf{598}, 390 (1996)

\bibitem{Sch03}
A.~Schwenk, B.~Friman, G.E. Brown, Nucl. Phys. A \textbf{713}, 191 (2003)

\bibitem{Mau14}
S.~Maurizio, J.W. Holt, P.~Finelli, Phys. Rev. C \textbf{90}, 044003 (2014)

\bibitem{Dri17}
C.~Drischler, T.~Kr\"uger, K.~Hebeler, A.~Schwenk, Phys. Rev. C \textbf{95},
  024302 (2017)

\bibitem{Wam93}
J.~Wambach, T.L. Ainsworth, D.~Pines, Nucl. Phys. A \textbf{555}, 128 (1993)

\bibitem{Sch96}
H.J. Schulze, J.~Cugnon, A.~Lejeune, M.~Baldo, U.~Lombardo, Phys. Lett. B
  \textbf{375}, 1 (1996)

\bibitem{Cao06}
L.G. Cao, U.~Lombardo, P.~Schuck, Phys. Rev. C \textbf{74}, 064301 (2006)

\bibitem{Ma08}
J.~Margueron, H.~Sagawa, K.~Hagino, Phys. Rev. C \textbf{77}, 054309 (2008)

\bibitem{Chen93}
J.M.C. Chen, J.W. Clark, R.D. Dav{\'{e}}, V.V. Khodel, Nucl. Phys. A
  \textbf{555}, 59 (1993)

\bibitem{Fab05}
A.~Fabrocini, S.~Fantoni, A.Y. Illarionov, K.E. Schmidt, Phys. Rev. Lett.
  \textbf{95}, 192501 (2005)

\bibitem{Fab08}
A.~Fabrocini, S.~Fantoni, A.Y. Illarionov, K.E. Schmidt, Nucl. Phys. A
  \textbf{803}, 137 (2008)

\bibitem{Mu05}
H.~M{\"{u}}ther, W.H. Dickhoff, Phys. Rev. C \textbf{72}, 054313 (2005)

\bibitem{Ding16}
D.~Ding, A.~Rios, H.~Dussan, W.H. Dickhoff, S.J. Witte, A.~Carbone, A.~Polls,
  Phys. Rev. C \textbf{94}, 025802 (2016)

\bibitem{Gan08}
S.~Gandolfi, A.Y. Illarionov, S.~Fantoni, F.~Pederiva, K.E. Schmidt, Phys. Rev.
  Lett. \textbf{101}, 132501 (2008)

\bibitem{GanC09}
S.~Gandolfi, A.Y. Illarionov, F.~Pederiva, K.E. Schmidt, S.~Fantoni, Phys. Rev.
  C \textbf{80}, 045802 (2009)

\bibitem{Abe09}
T.~{Abe}, R.~{Seki}, Phys. Rev. C \textbf{79}, 054003 (2009)

\bibitem{Gez08}
A.~Gezerlis, J.~Carlson, Phys. Rev. C \textbf{77}, 032801 (2008)

\bibitem{Gez10}
A.~Gezerlis, J.~Carlson, Phys. Rev. C \textbf{81}, 025803 (2010)

\bibitem{Krot80}
E.~Krotscheck, J.W. Clark, Nucl. Phys. A \textbf{333}, 77 (1980)

\bibitem{Krot81}
E.~Krotscheck, R.A. Smith, A.D. Jackson, Phys. Rev. B \textbf{24}, 6404 (1981)

\bibitem{Wir95}
R.B. Wiringa, V.G.J. Stoks, R.~Schiavilla, Phys. Rev. C \textbf{51}, 38 (1995)

\bibitem{Ho06}
C.J. Horowitz, A.~Schwenk, Phys. Lett. B \textbf{638}, 153 (2006)

\bibitem{Bo98}
G.H. Bordbar, M.~Modarres, Phys. Rev. C \textbf{57}, 714 (1998), and references
  therein

\bibitem{Bo97}
G.H. Bordbar, M.~Bigdeli, Phys. Rev. C \textbf{75}, 045804 (2007)

\bibitem{Mo15}
M.~Modarres, A.~Tafrihi, Nucl. Phys. A \textbf{941}, 212 (2015)

\bibitem{Fried81}
B.~Friedman, V.R. Pandharipande, Nucl. Phys. A \textbf{361}, 502 (1981)

\bibitem{Akmal98}
A.~Akmal, V.R. Pandharipande, D.G. Ravenhall, Phys. Rev. C \textbf{58}, 1804
  (1998)

\bibitem{Cl79}
J.W. Clark, L.R. Mead, E.~Krotscheck, K.E. K{\"{u}}rten, M.L. Ristig, Nucl.
  Phys. A \textbf{328}, 45 (1979)

\bibitem{Clark79}
J.W. Clark, Prog. Part. Nucl. Phys. \textbf{2}, 89 (1979)

\bibitem{Bomb05}
I.~Bombaci, A.~Fabrocini, A.~Polls, I.~Vida{\~{n}}a, Phys. Lett. B
  \textbf{609}, 232 (2005)

\bibitem{Marg07}
J.~Margueron, E.~van Dalen, C.~Fuchs, Phys. Rev. C \textbf{76}, 034309 (2007)

\bibitem{Baldo08}
M.~Baldo, C.~Maieron, Phys. Rev. C \textbf{77}, 015801 (2008)

\bibitem{Heb10}
K.~Hebeler, A.~Schwenk, Phys. Rev. C \textbf{82}, 014314 (2010)

\bibitem{Sch05}
A.~Schwenk, C.J. Pethick, Phys. Rev. Lett. \textbf{95}, 160401 (2005)

\bibitem{Tew16}
I.~Tews, S.~Gandolfi, A.~Gezerlis, A.~Schwenk, Phys. Rev. C \textbf{93}, 024305
  (2016)

\bibitem{Ep09}
E.~Epelbaum, H.~Krebs, D.~Lee, U.G. Mei{\ss}ner, Eur. Phys. J. A \textbf{40},
  199 (2009)

\bibitem{Carl03}
J.~Carlson, J.~Morales, V.R. Pandharipande, D.G. Ravenhall, Phys. Rev. C
  \textbf{68}, 025802 (2003)

\bibitem{Gan09}
S.~Gandolfi, A.Y. Illarionov, K.E. Schmidt, F.~Pederiva, S.~Fantoni, Phys. Rev.
  C \textbf{79}, 054005 (2009)

\bibitem{Lo11}
A.~Lovato, O.~Benhar, S.~Fantoni, A.Y. Illarionov, K.E. Schmidt, J. Phys. Conf.
  Ser. \textbf{336}, 012016 (2011)

\bibitem{Ben76}
O.~Benhar, C.C. {Degli Atti}, A.~Kallio, L.~Lantto, P.~Toropainen, Phys. Lett.
  B \textbf{60}, 129 (1976)

\bibitem{Pavlou09}
G.~Pavlou, Master's thesis, University of Athens (2009)

\bibitem{Wir02}
R.B. Wiringa, S.C. Pieper, Phys. Rev. Lett. \textbf{89}, 182501 (2002)

\bibitem{private}
E.~Krotscheck (2017), private communication

\bibitem{Ham13}
H.W. Hammer, A.~Nogga, A.~Schwenk, Rev. Mod. Phys. \textbf{85}, 197 (2013)

\bibitem{zuo}
W.~Zuo, Z.~Li, G.~Lu, J.~Li, W.~Scheid, U.~Lombardo, H.J. Schulze, C.~Shen,
  Phys. Lett. B \textbf{595}, 44 (2004)

\bibitem{zhou}
X.R. {Zhou}, G.F. {Burgio}, U.~{Lombardo}, H.J. {Schulze}, W.~{Zuo}, Phys. Rev.
  C \textbf{69}, 018801 (2004)

\bibitem{world}
J.W. Clark, \emph{{Fifty Years of Nuclear BCS \textnormal{Chapter 27, edited by
  R. A. Broglia and V. Zelevinsky}}} (World Scientific Publishing, 2013)

\end{thebibliography}
%
%
%

\end{document}